\documentclass[prl, showpacs, twocolumn, floatfix]{revtex4}
\usepackage{graphicx}
\usepackage{amsmath, amsfonts, amssymb, bm}
\begin{document}
\title{Measuring photon-photon interactions via photon detection}
\author{Mihai A. \surname{Macovei}}
\email{mihai.macovei@mpi-hd.mpg.de}
\affiliation{Max-Planck-Institut f\"{u}r Kernphysik, Saupfercheckweg
1, D-69117 Heidelberg, Germany}
\date{\today}
\begin{abstract}
The strong non-linearity plays a significant role in physics, particularly, 
in designing of novel quantum sources of light and matter as well as in 
quantum chemistry or quantum biology. In simple systems, the photon-photon 
interaction can be determined analytically. However, it becomes challenging 
to obtain it for more complex systems. Therefore, we show here how to measure 
strong non-linearities via allowing the sample to interact with a weakly 
pumped quantized leaking optical mode. We found that the detected mean-photon 
number versus pump-field frequency shows several peaks. Interestingly, the 
interval between neighbour peaks equals the photon-photon interaction potential. 
Furthermore, the system exhibits sub-Poissonian photon statistics, entanglement 
and photon switching with less than one photon. Finally, we connect our study 
with existing related experiments.
\end{abstract}
\maketitle
The non-linear interactions are perhaps the most investigated fundamental 
problems in modern physics and related subjects. An enormous amount of 
novel theoretical and experimental work has been carried out in this field. 
Significant results regarding this issue were obtained in quantum optics 
\cite{DW,IM,LK,SM,MX,NG,TLF,TA,TL,exp_BD,GK,ent_exp}, ultracold atomic 
gases \cite{HR,ZH,exp,exp_b,ph_b}, quantum electronics 
\cite{QN,GKN,MSK,GSA,PL,NMR,CQE,MB,MS,JBA} or quantum biochemistry \cite{BCH}. 
Particularly, strongly interacting single photons were discussed in \cite{IM} 
while quantum entanglement of ultraslow single photons and quantum switching 
were investigated in Ref.~\cite{LK}, respectively. Enhanced Kerr non-linearity 
was reported in \cite{SM,MX,NG,TLF}. Large Kerr non-linearity has possible 
applications in quantum non-demolition measurements and for quantum logic 
gates or for optical squeezing and studies of non-locality. The effect of Kerr 
non-linearity on the slow light propagation was discussed in \cite{TA} whereas 
turning light into a liquid via atomic coherence effects was studied in \cite{TL}. 
A procedure by which strong high-order non-linearities can be synthesized out of 
low-order non-linearities was proposed in \cite{exp_BD} while the quantum 
dissipative chaos in the statistics of excitation numbers was investigated in 
\cite{GK}. Entanglement via the Kerr non-linearity in an optical fiber system was 
experimentaly demonstrated in \cite{ent_exp}.

A number of experiments were performed emphasizing intresting non-linear 
phenomena in cold atomic samples \cite{HR,ZH,exp,exp_b} and trapped single-atom 
systems in optical cavities \cite{ph_b}. However, atomic 
systems are known to exhibit not so large non-linear couplings compared with 
the corresponding decay rates. Hence, a significant effort was devoted towards 
finding systems showing ultrahigh non-linearities. Remarkable, giant Kerr 
non-linearities were found in quantum nanosystems \cite{QN} and quantum circuit 
systems \cite{GKN}. The non-liniarities are further responsible for a number of 
fascinating phenomena. Bose-Einstein condensation, bistability and 
electromagnetic-field-induced transparency in high-density exciton systems were 
shown to occur in \cite{MSK,GSA}, while polariton quantum blockade in a photonic 
dot was investigated in Ref.~\cite{PL}. Signatures for a classical to quantum 
transition of a driven non-linear nanomechanical resonator and photon-number 
squeezing in circuit quantum electro\-dynamics were studied in \cite{NMR} and 
\cite{CQE}, respectively. A proposal for detecting single-phonon transitions in 
a single nanoelectromechanical system was discussed as well \cite{MB}. Other 
non-linear effects involving macroscopic systems, superconducting qubits or 
biochemical samples were reported in \cite{MS,JBA,BCH} further advancing the 
research in these areas. 

In many of the above mentioned effects, at least some knowledge on the non-linear 
interactions is required. A number of simpler samples allow the non-linearity 
to be determined even analytically. However, it becomes difficult to obtain it 
for more sophisticated systems. Therefore, here, we show how to 
measure the third-order non-linear susceptibility $\chi^{(3)}$ which is responsible 
for the boson-boson interaction potentials, for instance. We model the system as a 
pumped non-linear quantum oscillator. The examined sample additionally interacts with 
a quantized lossy field mode. The applied field is weak such that the non-linearity 
is not affected by the external driving. We found that in the long-time limit, the 
detected photons show several peaks as function of external field detuning when the 
photon-photon non-linearity is larger than the quantized mode damping. Remarkably, 
the frequency intervals between the neighbour peaks are equal with the photon-photon 
interaction potential allowing us to determine it. As the strength of the pumping 
field is increased, the observed mean photon number shows strong asymmetrical behaviors. 
At lower pumping intensities switching with less than one photon occurs. Moreover, the 
switching effect improves for even larger non-linear interactions. The photon statistics 
can be sub-Poissonian. These results would disappear for vanishing boson-boson interactions. 
Note that for particular parameters one can create an entangled single-photon state 
$|\Psi_{\pm}\rangle=\bigl (|0\rangle \pm |1\rangle\bigr)/\sqrt{2}$ with a high 
fidelity. Here, $|0\rangle$ and $|1\rangle$ are the number state basis with zero 
and one photon, respectively. Therefore, the study is of significance for applications 
ranging from optical communication to quantum processing of information.

We proceed by briefly introducing the main steps of the analytical formalism involved 
and then rigorously describing the obtained results. The Hamiltonian characterizing 
the interaction of a non-linear oscillator possessing the frequency $\omega$ with a 
coherent source of frequency $\omega_{L}$, in a frame rotating at $\omega_{L}$, is:
\begin{eqnarray}
H=-\Delta a^{\dagger}a \mp \alpha a^{\dagger^2}a^{2} + \epsilon(a^{\dagger}+a), 
\label{Hm}
\end{eqnarray}
where $\Delta=\omega_{L}-\omega$ and $\alpha >0$ signifies the photon-photon 
non-linearity ($\alpha \propto {\rm Re}[\chi^{(3)}]$), while $\epsilon$ is 
the amplitude of the coherent driving. $a^{\dagger}$ and $a$ are the creation 
and the annihilation operator of the quantum oscillator, respectively, and 
obeying the standard bosonic commutation relations, i.e., $[a,a^{\dagger}]=1$, 
and $[a,a]=[a^{\dagger},a^{\dagger}]=0$. The $\mp$ sign in Eq.~(\ref{Hm}) 
accounts for an attractive or repulsive boson-boson interaction, respectively. 
Actually, the Hamiltonian (\ref{Hm}) is abundantly investigated and characterizes 
a wide range of processes. A formidable approach to analyze the driven damped 
non-linear oscillator is via the master equation. Hence, the system is described 
by the reduced density operator, which in the interaction picture and under the 
usually applicable Born-Markov and rotating-wave approximations satisfies the 
master equation \cite{HM,WM}:
\begin{eqnarray}
\dot \rho = -i[H,\rho] - \kappa\bigl ([a^{\dagger},a\rho] + [\rho a^{\dagger},a] \bigr),
\label{ME}
\end{eqnarray}
where $\kappa$ is the decay rate of the quantized optical oscillator mode 
and the overdot denotes differentiation with respect to time.

In order to investigate Eq.~(\ref{ME}) in the long-time limit and for 
weak excitations in more details, we first shall apply the Holstein-Primakoff 
transformations \cite{HP}, and then follow the solving procedure developed in 
\cite{kl} for two-level particles. We define a state $|q\rangle$ denoting $N-q$ 
spins pointing up and $q$ spins down. The raising and lowering spin operators 
act on this state as follows: $S^{+}|q\rangle=\sqrt{q(2s-q+1)}|q-1\rangle$ and 
$S^{-}|q\rangle=\sqrt{(q+1)(2s-q)}|q+1\rangle$, where $2s$ gives the number of 
excitations (spins) in the system, and $0 \le q \le2s$. The spin operators satisfy 
the usual commutation relations of su(2) algebra, i.e., $[S_{z},S^{\pm}]=\pm S^{\pm}$ 
and $[S^{+},S^{-}]=2S_{z}$, where $S_{z}|q\rangle=(s-q)|q\rangle$. Taking into account 
these properties, one can deduce that \cite{HP}:
\begin{eqnarray}
S^{+}=\sqrt{2s}\sqrt{1-\frac{a^{\dagger}a}{2s}}a, &~~& S^{-}=\sqrt{2s}a^{\dagger}
\sqrt{1-\frac{a^{\dagger}a}{2s}}, \nonumber \\
S_{z} &=& s-a^{\dagger}a. \label{HP}
\end{eqnarray}
Substituting (\ref{HP}) in the steady-state form of Eq.~(\ref{ME}), 
observing that $S^{-}S^{+}=2sa^{\dagger}a-a^{\dagger^2}a^{2}$, and 
assuming lower excitations, i.e. $a^{\dagger}a/2s \ll 1$, one arrives 
at the following long-time solution for the normalized diagonal 
elements $P_{q}= \langle q|\rho_{s}|q\rangle$ of the density matrix $\rho_{s}$:
\begin{eqnarray}
P_{q} = Z^{-1}\sum^{2s-q}_{m=0}C_{mm}\frac{(q+m)!(2s-q)!}{q!(2s-q-m)!}. 
\label{Pq} 
\end{eqnarray}
Here, the normalization constant $Z$ is given by the expression:
\begin{eqnarray}
Z = \sum^{2s}_{m=0}C_{mm}\frac{(2s+m+1)!(m!)^{2}}{(2s-m)!(2m+1)!}, 
\label{Z}
\end{eqnarray}
where $C_{mm}=|\sigma|^{-2m}|\Gamma(1+m+i\phi^{*})/[m!\Gamma(1+i\phi^{*})]|^{2}$ 
and $\sigma = \Omega/(\gamma \pm i\alpha)$, $\phi=\pm(2s\alpha \pm \Delta)
/(\gamma \pm i\alpha)$, while $\Omega=\epsilon/\sqrt{2s}$ and $\gamma=\kappa/2s$. 
With the help of the steady-state solution (\ref{Pq}), one can calculate the 
oscillator's variables of interest. In particular, the mean number of excitations 
in the system $\langle n\rangle \equiv \langle a^{\dagger}a\rangle$ as well as 
its second-order correlation function $G^{(2)}(0)\equiv \langle a^{\dagger^2}a^{2}\rangle $ 
can be obtained from the folowing expressions:
\begin{eqnarray}
\langle n\rangle=\sum^{2s}_{q=0}qP_{q}, ~~
G^{(2)}(0)=\sum^{2s}_{q=0}q(q-1)P_{q}. \label{nm}
\end{eqnarray}
Because we considered the weak excitation limit, the solution (\ref{Pq}) is 
valid as long as $\Omega \ll \alpha$. This means that the amplitude of the 
external driving $\epsilon$ should be less or of the order of boson-boson 
potential $\alpha$. Thus, our approach is well suitable to describe low-photon 
processes with mean-photon numbers of the order of few photons.
\begin{figure}[t]
\includegraphics[width=8.6cm]{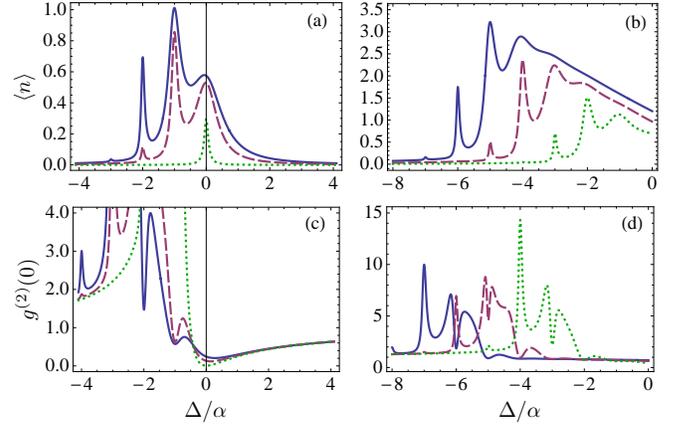}
\caption{\label{fig-1}(color online) The mean photon number $\langle n\rangle$ 
as well as the normalized second-order correlation function $g^{(2)}(0)$ as 
function of scaled detuning $\Delta/\alpha$. Here (a,c) $\Omega/\alpha=0.06, 0.04, 0.006$ 
while (b,d) $\Omega/\alpha=0.3, 0.2, 0.1$ for the solid, the dashed and the dotted curves, 
respectively. Other parameters are: $\gamma/\alpha=10^{-3}$ and $2s=50$.}
\end{figure}

We further focus on investigating the properties of the driven and damped 
non-linear oscillator using Eqs.~(\ref{Pq}-\ref{nm}). As a first result, 
Fig.~(\ref{fig-1}) shows the steady-state dependence of the mean photon 
number $\langle n\rangle$ and the normalized second-order correlation function 
$g^{(2)}(0)=G^{(2)}(0)/\langle n\rangle^{2}$ versus the external field detuning 
$\Delta/\alpha$ for the negative sign in Eq.~(\ref{Hm}). There, (a,c) describe 
the weak field limit whereas (b,d) depict pumping with a moderate coherent source. 
Interestingly, single-photon light with sub-Poissonian photon statistics can be 
obtained (see Fig.~{\ref{fig-1}}(a,c) near $\Delta/\alpha=-1$). This is not trivial 
since even weaker fields lead to super-Poissonian photon statistics 
(see Fig.~{\ref{fig-1}}(a,c) at $\Delta/\alpha < -1$). However, the two-photon 
correlator $G^{(2)}(0)$ has small values here and larger values for $g^{(2)}(0)$ 
are due to even smaller denominator, i.e., $\langle n\rangle^{2}$. For stronger 
driving, super-Poissonian photon statistics almost always occurs for 
$\Delta/\alpha < -1.5$ and Poissonian or sub-Poissonian statistics for 
$\Delta/\alpha > -1.5$ (see Fig.~{\ref{fig-1}}(b,d)). Furthermore, the mean photon 
number shows strong asymmetrical behaviors. The asymmetry is more pronounced for 
intenser external fields (compare Fig.~{\ref{fig-1}}(a) and Fig.~{\ref{fig-1}}(b)). 
Notably, the multi-peak behaviors of the mean photon number can help us to determine 
the non-linearity $\alpha$. One can observe that the scaled frequency interval between 
the neighbour peaks in Fig.~\ref{fig-1}(a,b) equals unity and, thus, the corresponding 
inter-peak frequency will be equal to the non-linearity $\alpha$. Therefore, the 
boson-boson non-linearity can be extracted by measuring the mean photon number against 
the external pumping field detuning. The only condition is that the non-linearity 
$\alpha$ should be at least few times larger than the damping of the quantized mode 
$\kappa$. This is well-satisfied in a wide range of systems \cite{GKN}. Note that 
the peaks at $\Delta/\alpha=-6$ in Fig.~(\ref{fig-1}b) and Fig.~(\ref{fig-1}d) are 
slightly smaller than those obtained numerically. The inter-peaks interval is obtained 
correctly for even stronger pumping, however, the magnitude of the peaks as well as 
their positions will be given only aproximately. The same results occur for the 
positive sign in Eq.~(\ref{Hm}), i.e., for repulsive boson-boson interactions, though 
the corresponding curves in Fig.~(\ref{fig-1}) will be mirrored with respect to the 
$0Y$ axis. Therefore, for simplicity, we shall consider further only attractive 
boson-boson interactions.

An explanation of the multi-peak behaviors in Fig.~(\ref{fig-1}) can be found by 
representing the Eq.~(\ref{ME}) via the number state basis, i.e. 
$P_{mn}=\langle m|\rho_{s}|n\rangle$. Then, for the negative sign in Eq.~(\ref{Hm}), 
one arrives at the following terms next to $P_{mn}$, that is, 
$P_{mn}[i(m-n)\{\Delta+\alpha(m+n-1)\}-\kappa(m+n)]$. One can observe here that 
resonances occur at $\Delta+\alpha(m+n-1)=0$, when $\alpha > \kappa$. The off-diagonal 
elements induced by the driving field in the steady-state play a crucial role here. 
In their absence the mean photon number, for instance, would be insensitive on 
photon-photon interactions. In a recent work \cite{ferm}, additional resonances were 
found which may help to extract the non-linear parameters, though in a more complicated 
setup.

\begin{figure}[t]
\includegraphics[width=8.6cm]{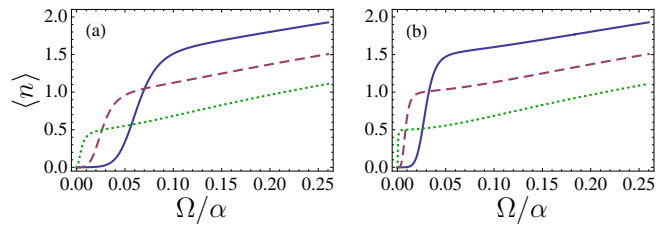}
\caption{\label{fig-2}(color online) The mean photon number $\langle n\rangle$ 
versus the intensity of the pumping field $\Omega/\alpha$. Solid line is for 
$\Delta/\alpha=-2$, dashed curve corresponds to $\Delta/\alpha=-1$ while the 
dotted line to $\Delta/\alpha=0$. Here, (a) $\gamma/\alpha=10^{-3}$ and (b) 
$\gamma/\alpha=10^{-4}$ with $2s=50$.}
\end{figure}

To further describe our system, we plot in Fig.~(\ref{fig-2}) the dependence 
of the non-linear oscillator's mean photon number as function of the applied 
field intensity. The detunings are adjusted to values giving the peaks in 
Fig.~(\ref{fig-1}a). Switching with less than one photon is observed here 
(see the dotted line in Fig.~\ref{fig-2}). The mean photon number abruptly 
increases as the pumping strength varies only a little. For instance, the 
transition among states with 
$\langle n\rangle =0$ to $\langle n\rangle \approx 1$ occurs while the intensity 
of the external pumping field changes between $\Omega/\alpha \approx 0.005$ and 
$\Omega/\alpha \approx 0.02$ (see the dashed curve in Fig.~\ref{fig-2}b). The 
switching phenomenon considerably improves for larger boson-boson interactions 
(compare Fig.~\ref{fig-2}a and Fig.~\ref{fig-2}b). Positive values of $\Delta$, 
away from resonance, do not lead to switching effects. These critical behaviors take 
place for repulsive photon-photon interactions too, but for $\Delta \ge 0$. Thus, the 
appropriately prepared non-linear sample can be presented as a single-photon optical 
switching devise. Finally, for comparison, we have obtained the oscillator's mean 
photon number and its second-order correlation function without taking into account 
of the boson-boson interaction potential, i.e., we set $\alpha=0$: 
\begin{eqnarray}
\langle n \rangle =\frac{\epsilon^{2}}{\kappa^{2}+\Delta^{2}}, ~~~  
G^{(2)}(0) =\frac{\epsilon^{4}}{(\kappa^{2}+\Delta^{2})^{2}}. \label{nK}
\end{eqnarray}
In this case, however, the rich variety of effects shown in Fig.~(\ref{fig-1}) 
and Fig.~(\ref{fig-2}) vanishes and the photon statistics allways shows Poissonian 
photon distribution.

In the experiment described in Ref.~\cite{exp}, few-photon switching, Kerr 
non-linearity and dispersive optical bistability of a Fabry-Perot optical 
resonator due to the displacement of ultracold atoms trapped within were 
reported. There, the photon asymmetry similar to the one shown in 
Fig.~{\ref{fig-1}(a,b)} was observed as well (see also Fig.~2 in \cite{exp}). 
Though experimentally, they got multiple peaks of the intracavity mean photon 
number their nature was not explained. Based on our results, we can conjecture 
that the interval between the main peaks around the central one of the 
intracavity mean photon number obtained in \cite{exp} may give the induced 
boson-boson non-linearity. Optical switching with single photons was observed 
as well \cite{sosw}.

\begin{figure}[t]
\includegraphics[width=6cm]{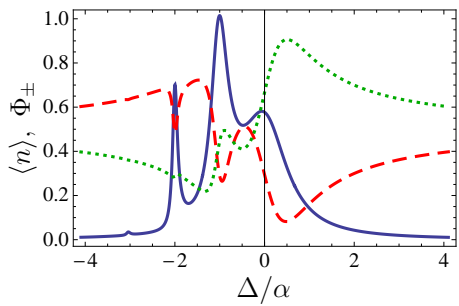}
\caption{\label{fig-3}(color online) The mean photon number $\langle n\rangle$ 
as well as the fidelities $\Phi_{\pm}$ against the pumping field detuning 
$\Delta/\alpha$. The solid line depicts the mean photon number $\langle n\rangle$ 
while the dashed and the dotted curves correspond to $\Phi_{-}$ and $\Phi_{+}$, 
respectively. Here, $\Omega/\alpha=0.06$, $\gamma/\alpha=10^{-3}$ and $2s=50$.}
\end{figure}

In order to look at more details in the few-photon processes emphasized here, in 
Fig.~(\ref{fig-3}), we plot the fidelities 
$\Phi_{\pm}=\langle \Psi_{\pm}|\rho_{s}|\Psi_{\pm}\rangle$ of entangled states 
\begin{eqnarray}
|\Psi_{\pm}\rangle=\frac{1}{\sqrt{2}}\bigl (|0\rangle \pm |1\rangle\bigr),
\label{ent}
\end{eqnarray}
where $|0\rangle$ and $|1\rangle$ are the number state basis with zero and 
one photon, respectively. There, the dashed curve corresponds to $\Phi_{-}$ 
while the dotted one to $\Phi_{+}$. The solid line shows the non-linear 
oscillator's mean photon number $\langle n\rangle$. The parameters are 
the same as for the solid line in Fig.~(\ref{fig-1}a). One can observe 
here that at $\Delta/\alpha=0.5$ the entangled state $|\Psi_{+}\rangle = 
\bigl(|0\rangle + |1\rangle \bigr)/\sqrt{2}$ is created with a fidelity 
$\Phi_{+} \approx 0.9$ and $\langle n\rangle \approx 0.4$, while the photon 
statistics is sub-Poissonian (see Fig.~\ref{fig-1}c). On the other side, at 
$\Delta/\alpha=-1.5$ the entangled state $|\Psi_{-}\rangle = \bigl(|0\rangle 
- |1\rangle \bigr)/\sqrt{2}$ occurs with a fidelity $\Phi_{-} \approx 0.7$ 
and $\langle n\rangle \approx 0.2$, and where photon statistics is 
super-Poissonian (see Fig.~\ref{fig-1}c). Larger fidelities mean establishing 
of quantum coherences in the system and, thus, proving the quantum nature of 
the states (\ref{ent}). Note that for detunings approximately within 
$-2.1 < \Delta/\alpha < 0$ one has $\Phi_{+}+\Phi_{-}<1$ denoting the existence 
of higher Fock states, i.e. $|n>1\rangle$, though with smaller probabilities 
(see Fig.~\ref{fig-3}). Therefore, we have demonstrated here switching and 
entanglement with less than one photon processes. 

In summary, we have shown how to measure the photon-photon interaction potential 
via weakly pumping the quantized mode with which the examined system interacts. 
The non-linear effect induces multiple peaks in the mean photon number steady-state 
behaviors with inter-peaks intervals equating the required non-linearity. For this 
to occur the boson-boson non-linearity should be larger than the quantized mode 
damping. This requirement is well-fulfilled in a wide range of systems. Due to the 
involved non-linearity, an asymmetry in the first- and second-order field correlation 
functions was observed as well. In addition, we have demonstrated switching and 
high-fidelity entanglement with less than one photon process. Sub-Poissonian photon 
statistics occurs for particular external controlable parameters. The reported results 
can be tested with existing experiments.

\end{document}